\pdfoutput=1

\documentclass[aps,twocolumn,amsmath,amssymb,preprintnumbers,floatfix,prl,superscriptaddress,longbibliography]{revtex4-1}

\usepackage[utf8]{inputenc}
\usepackage{newtxtext}
\usepackage[upint]{newtxmath}
\usepackage{microtype}
\usepackage{textcomp}
\usepackage{eucal}
\usepackage{bm}
\usepackage{siunitx}
\usepackage{comment}

\usepackage{enumerate}
\usepackage{amsfonts}
\usepackage{amsmath}
\usepackage{amssymb}
\usepackage{color}
\usepackage{soul}

\usepackage{graphicx}

\usepackage[colorlinks,allcolors=blue]{hyperref}
\usepackage[capitalize]{cleveref}

\definecolor{DarkRed}{rgb}{0.65,0,0}%
\definecolor{Green}{rgb}{0,0.3,0.3}
\definecolor{Purple}{rgb}{0.3,0,0.65}
\definecolor{Red}{rgb}{1,0,0}
\definecolor{Blue}{rgb}{0,0,0.85}
\definecolor{Magenta}{rgb}{1,0,1}

\newcommand{\ve}[1]{\boldsymbol{#1}}

\newcommand{\eg}{\textit{e.g. }}

\newcommand{\be}{\begin{equation}}
\newcommand{\ee}{\end{equation}}

\newcommand{\prlsection}[1]{\textit{#1}.\kern0.05em---\kern0.05em\ignorespaces}

\begin{document}
\title{Magnetic control of superconducting heterostructures using compensated antiferromagnets}
\author{Lina G. Johnsen}
\email[Corresponding author: ]{lina.g.johnsen@ntnu.no}
\affiliation{Center for Quantum Spintronics, Department of Physics, Norwegian \\ University of Science and Technology (NTNU), NO-7491 Trondheim, Norway}
\affiliation{These authors contributed equally to this work}
\author{Sol H. Jacobsen}
\affiliation{Center for Quantum Spintronics, Department of Physics, Norwegian \\ University of Science and Technology (NTNU), NO-7491 Trondheim, Norway}
\affiliation{These authors contributed equally to this work}
\author{Jacob Linder}
\affiliation{Center for Quantum Spintronics, Department of Physics, Norwegian \\ University of Science and Technology (NTNU), NO-7491 Trondheim, Norway}

\begin{abstract}

Due to the lack of a net magnetization both at the interface and in the bulk, antiferromagnets with compensated interfaces may appear incapable of influencing the phase transition in an adjacent superconductor via the spin degree of freedom. We here demonstrate that such an assertion is incorrect by showing that proximity-coupling a compensated antiferromagnetic layer to a superconductor--ferromagnet heterostructure introduces the possibility of controlling the superconducting phase transition. The superconducting critical temperature can in fact be modulated by rotating the magnetization of the single ferromagnetic layer within the plane of the interface, although the system is invariant under rotations of the magnetization in the absence of the antiferromagnetic layer. 
Moreover, we predict that the superconducting phase transition can trigger a reorientation of the ground state magnetization.
Our results show that a compensated antiferromagnetic interface is in fact able to distinguish between different spin-polarizations of triplet Cooper pairs.

\end{abstract}
\maketitle

\begin{figure}[b]
    \centering
    \includegraphics[width=\columnwidth]{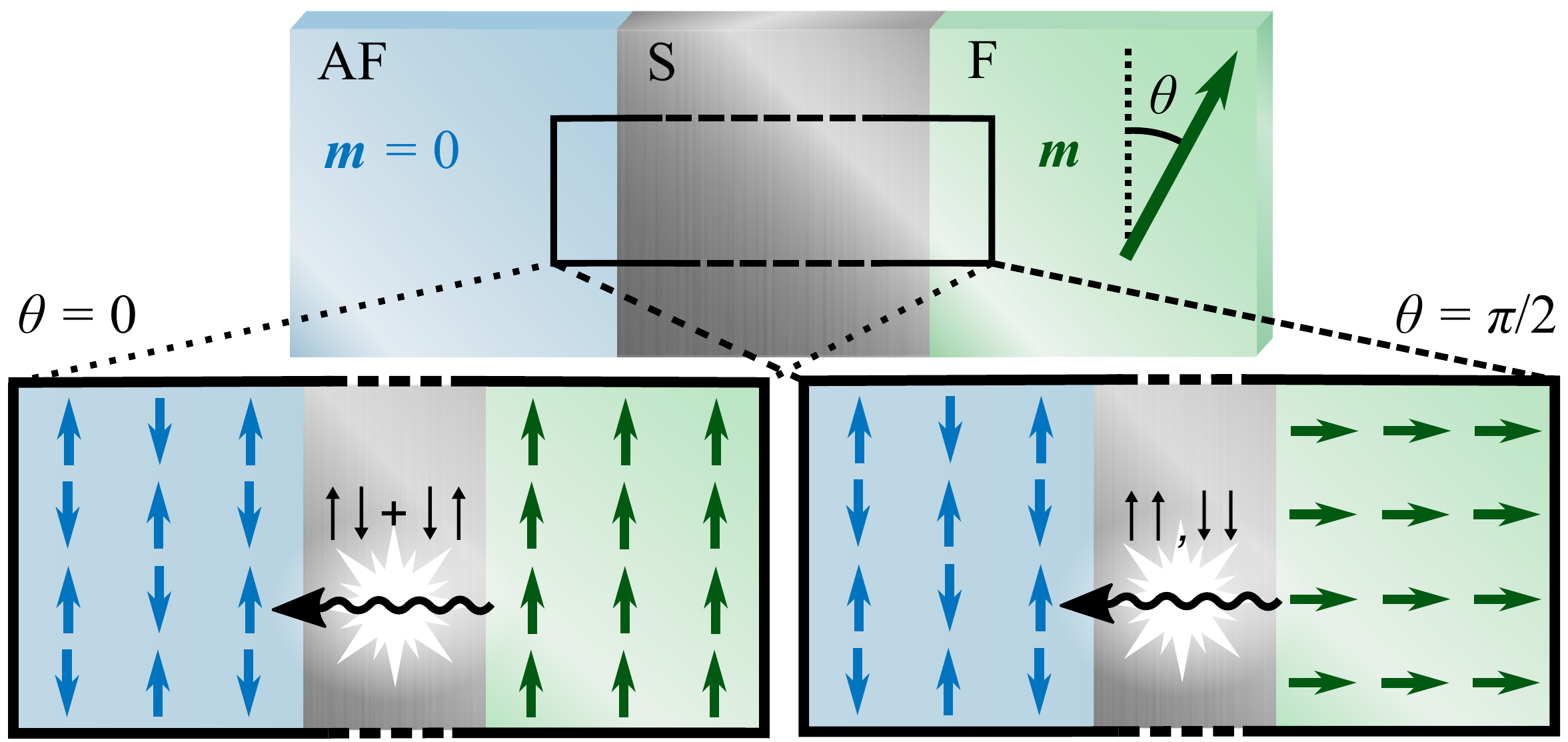}
    \caption{Although the total magnetization $\ve{m}$ of the antiferromagnet is zero (top), the interaction between the local magnetic moments of the antiferromagnet and ferromagnet affects the generation of spin-triplet Cooper pairs. When the magnetic moments are aligned (left), only opposite-spin triplets are present. When they are misaligned, the opposite-spin triplets created at the ferromagnetic interface are partially, or entirely in the perpendicular case (right), rotated into equal-spin triplets relative to the magnetic moments in the antiferromagnet. These triplets scatter differently at the antiferromagnetic interface compared to the triplets that exist in the parallel case, resulting in a weakened singlet condensate. Since the triplet generation only depends on the misalignment between the magnetic moments, we can choose to control the triplet channels by rotating the magnetization within the plane of the interface.
	}
    \label{fig:illustration}
\end{figure}

\textit{Introduction}.--- 
Proximity effects in heterostructures consisting of ferromagnets (F) and conventional superconductors (S) have been widely studied, in part due to the possibility of creating spin-polarized Cooper pairs \cite{golubov_rmp_04,buzdin_rmp_05,bergeret_rmp_05,izyumov_pu_02}. When the spin-singlet Cooper pairs of a conventional superconductor enters a ferromagnetic material , the spin-splitting of the energy bands of the ferromagnet gives rise to opposite-spin triplets as spin-up and spin-down electrons acquire different phases upon propagation.  Further, the opposite-spin triplets can be rotated into equal-spin triplets with respect to a ferromagnet with a differently oriented magnetization \cite{eschrig_pt_11, eschrig_rpp_15}. Such triplets can also exist in structures with a single inhomogeneous ferromagnet \cite{bergeret_prl_01_b,kadigrobov_epl_01,wu_prl_12,chiodi_epl_13}. Equal-spin triplets relative the magnetization direction are more robust to pair-breaking effects from the ferromagnetic exchange field. The generation of equal-spin triplets therefore causes an increased leakage of Cooper pairs from the superconducting region and a weakening of the superconducting condensate. By controlling the singlet to triplet conversion, we can thus manipulate the superconducting condensation energy and the critical temperature \cite{buzdin_jetp_90,radovic_prb_91,buzdin_jetp_91,jiang_prl_95,mercaldo_prl_96}.

The singlet to triplet generation can be controlled by adjusting the misalignment between  two ferromagnets proximity-coupled to a superconductor \cite{eschrig_pt_11, eschrig_rpp_15}. However, when combining these into F/S/F structures, the dominant effect on the superconducting condensation energy and the critical temperature $T_c$ is not the opening of the equal-spin triplet channels. Instead, the mutual compensation of the ferromagnetic exchange fields favour antiparallel alignment of the ferromagnets in order to minimize the field inside the superconductor \cite{gu_prl_02, moraru_prl_06, banerjee_natcom_14,gu_prl_15}. By arranging the materials in a S/F/F structure this effect becomes less prominent \cite{bergeret_prl_01,fominov_jetp_10,wu_prb_12_b,leksin_prl_12, jara_prb_14, wang_prb_14}, however this necessitates the ability to tune the orientations of the ferromagnets independently. It is therefore desirable to reduce the number of magnetic elements required to tune $T_c$ in order to minimize the stray field of the heterostructure. Stray fields would be a disturbance to neighboring elements if the heterostructure was part of a larger device architecture.

Previous studies have suggested introducing heavy-metal layers boosting the interfacial Rashba spin-orbit coupling in a S/F bilayer \cite{jacobsen_prb_15, ouassou_sr_16, simensen_prb_18, banerjee_prb_18, johnsen_prb_19, eskilt_prb_19, gonzalez-ruano_prb_20}. The Rashba spin-orbit field introduces additional symmetry breaking \cite{johnsen_prl_20} that allows for control over the spin-triplet channels when rotating the magnetization of the single ferromagnetic layer. However, for a structure with purely Rashba spin-orbit coupling, a variation in the triplet generation for in-plane rotations of the magnetization is only possible in ballistic-limit systems \cite{johnsen_prb_19}, while additional Dresselhaus spin-orbit coupling is needed for such an in-plane effect in the diffusive limit \cite{jacobsen_prb_15}. 
In this work, we consider another possibility for controlling the spin-triplet channels, namely replacing one of the ferromagnetic layers in the F/S/F structure with an antiferromagnet with a compensated interface.

Antiferromagnets (AF) provide a magnetic structure with zero net magnetization \cite{baltz_rmp_14}. When proximity-coupling antiferromagnets to other materials, antiferromagnets therefore have the advantage of not emitting an external field to its surroundings. We therefore avoid vortex formation and demagnetizing currents in adjacent superconductors, and the magnetization of an adjacent ferromagnet can be easily controlled. Also, the magnetic moments of the antiferromagnet are insensitive to disturbing magnetic fields \cite{jungwirth_nn_16}.
Studies of quasiparticle reflection \cite{bobkova_prl_05,andersen_prb_05}, Josephson effects \cite{bell_prb_03, andersen_prl_06,  komissinskiy_prl_07,enoksen_prb_13, bulaevskii_prb_17,zhou_epl_19, rabinovich_prr_19}, the superconducting critical temperature \cite{hubener_jpcm_02, westerholt_prl_05, robinson_arxiv_08,wu_apl_13} and the critical field \cite{wu_apl_13} in uncompensated superconductor--antiferromagnet structures have proven antiferromagnets to be applicable for manipulating the superconducting state, despite their zero net magnetization. It has also been shown that uncompensated antiferromagnetic insulators can induce spin-splitting in an adjacent superconductor \cite{kamra_prl_18}. 
Antiferromagnet--ferromagnet structures have shown interesting properties for spintronics applications, \eg magnetization switching mediated by spin-orbit torques \cite{fukami_nm_16,wang_sc_19}.

The above-mentioned works have mostly focused on uncompensated antiferromagnetic interfaces where there is an effective magnetization at the interface. Compensated antiferromagnetic interfaces have been claimed to be spin-inactive in several recent works, and only a few have reported a nonzero effect on an adjacent superconducting condensate \cite{bobkova_prl_05,andersen_prb_05}. We here consider a heterostructure consisting of a homogeneous ferromagnet, a conventional superconductor, and a compensated antiferromagnetic insulator.
We demonstrate that despite the zero net magnetization in the antiferromagnet, the misalignment between the magnetic moments of the antiferromagnet and ferromagnet allows for control over the spin-triplet amplitude, as illustrated in Fig.~\ref{fig:illustration}. 
This makes it possible to manipulate the superconducting phase transition by rotation of the ferromagnetic magnetization. 
Moreover, we predict that the suppression of the superconducting gap for misaligned magnetic moments leads to a modulation of the effective ferromagnetic anisotropy, potentially causing a magnetization reorientation driven by the superconducting phase transition. 
To the best of our knowledge, this manuscript presents the first prediction of a compensated antiferromagnetic interface being able to distinguish between different spin-polarizations of triplet Cooper pairs. The $T_c$ variation and magnetization reorientation predicted in our work is a direct manifestation of this new physical effect.

\textit{Theoretical framework}.---
We describe the AF/S/F heterostructure by the tight-binding Bogoliubov--de Gennes Hamiltonian
\begin{equation}
    \begin{split}
        H=&-t\sum_{\left<\ve{i},\ve{j}\right>,\sigma}c_{\ve{i},\sigma}^{\dagger}c_{\ve{j},\sigma}
        -\sum_{\ve{i},\sigma}\mu_{\ve{i}}c_{\ve{i},\sigma}^{\dagger}c_{\ve{i},\sigma}
        +\sum_{\ve{i}\in \text{AF}}V_{\ve{i}}n_{\ve{i},\uparrow}n_{\ve{i},\downarrow}\\
        &-\sum_{\ve{i}\in \text{S}}U_{\ve{i}}n_{\ve{i},\uparrow}n_{\ve{i},\downarrow}
        +\sum_{\ve{i}\in \text{F},\sigma,\sigma'}c_{\ve{i},\sigma}^{\dagger}\left(\ve{h}_{\ve{i}}\cdot\ve{\sigma}\right)_{\sigma,\sigma'}c_{\ve{i},\sigma'}.
    \end{split}
\end{equation}
The first two terms are present throughout the whole structure as they include nearest-neighbor hopping and the chemical potential. Above, $t$ is the hopping integral, $\mu_{\ve{i}}$ is the chemical potential at lattice site $\ve{i}$, and $c_{\ve{i},\sigma}^{\dagger}$ and $c_{\ve{i},\sigma}$ are the electron creation and annihilation operators at lattice site $\ve{i}$ for electrons with spin $\sigma$. 
The remaining three terms are only nonzero in their respective regions. In these terms, $V_{\ve{i}}>0$ is the on-site Coulomb repulsion giving rise to antiferromagnetism, $U_{\ve{i}}>0$ is the attractive on-site interaction giving rise to superconductivity, $\ve{h}_{\ve{i}}$ is the local magnetic exchange field giving rise to ferromagnetism, $n_{\ve{i},\sigma}\equiv c_{\ve{i},\sigma}^{\dagger}c_{\ve{i},\sigma}$ is the number operator, and $\ve{\sigma}$ is the vector of Pauli matrices. We choose the chemical potential in the antiferromagnetic region to be approximately zero so that the antiferromagnet behaves as an insulator. Throughout this work, all energies are scaled by the hopping integral $t$, and all length scales are scaled by the lattice constant. For simplicity, we set the Boltzmann and reduced Planck constants to one.

Our theoretical framework is well suited for describing heterostructures consisting of atomically thin layers in the ballistic limit, and it fully accounts for the crystal structure of the system. In our theoretical framework, the particular lattice geometry chosen (square) is not important to understand the triplet generation in the AF/S/F hybrid. Adding disorder and interfacial barriers would influence the magnitude of the predicted $T_c$-change, but not its existence.
For simplicity, we therefore consider a 2D square lattice of size $N_x \times N_y$ with interface normal along the $x$ axis. We assume that the ferromagnetic exchange field is oriented within the plane of the interface and that it is constant throughout the ferromagnetic layer. We describe the ferromagnetic exchange field as $\ve{h}=h[0,\sin(\theta),\cos(\theta)]$ in terms of the polar angle $\theta$ with respect to the $z$ axis.

The antiferromagnetic contribution is treated by a mean-field approach that preserves the spin-rotational invariance of the antiferromagnetic order parameter $\ve{M}_{\ve{i}}\equiv 4V_{\ve{i}}\left<\ve{S}_{\ve{i}}\right>/3$ \cite{andersen_thesis_04}. We write the antiferromagnetic term in the Hamiltonian in terms of the spin operator $\ve{S}_{\ve{i}}\equiv \frac{1}{2}\sum_{\sigma,\sigma'}c_{\ve{i},\sigma}^{\dagger}\ve{\sigma}_{\sigma,\sigma'}c_{\ve{i},\sigma'}$ and assume that the spin operator only weakly fluctuates around its expectation value so that $\ve{S}_{\ve{i}}=\left<\ve{S}_{\ve{i}}\right>+\delta_{\text{AF}}$. We neglect second order terms in the spin fluctuations $\delta_{\text{AF}}$.
The superconducting contribution is also treated by a mean-field approach where we similarly write $c_{\ve{i},\uparrow}c_{\ve{i},\downarrow}=\left<c_{\ve{i},\uparrow}c_{\ve{i},\downarrow}\right>+\delta_{\text{S}}$ and neglect second order terms in the fluctuations $\delta_{\text{S}}$. The superconducting gap $\Delta_{\ve{i}}\equiv U_{\ve{i}}\left<c_{\ve{i},\uparrow}c_{\ve{i},\downarrow}\right>$ is treated self-consistently.
We assume that the order parameter of the antiferromagnet is large compared to the superconducting gap, so that it is robust under reorientations of the magnetization of the ferromagnet. Under these assumptions it is not necessary to treat the antiferromagnetic order parameter self-consistently, and we assume it to have a constant absolute value $M$ and opposite signs on neighboring lattice sites. By solving both $\ve{M}_{\ve{i}}$ and $\Delta_{\ve{i}}$ self-consistently, we have verified that $\ve{M}_{\ve{i}}$ remains unchanged as $\ve{h}$ is rotated, although $\Delta_{\ve{i}}$ changes significantly. 

We diagonalize the Hamiltonian numerically by assuming periodic boundary conditions in the $y$ direction as outlined in the Supplemental Material \cite{Supplemental}, and calculate the spin-triplet amplitudes, the superconducting critical temperature $T_c$, and the free energy of the system.
The $s$-wave odd-frequency opposite- and equal-spin  triplet amplitudes are defined as
$S_{0,\ve{i}}(\tau)\equiv\sum_{\sigma}\left<c_{\ve{i},\sigma}(\tau)c_{\ve{i},-\sigma}(0)\right>$, and
$S_{\sigma,\ve{i}}(\tau)\equiv\left<c_{\ve{i},\sigma}(\tau)c_{\ve{i},\sigma}(0)\right>$,
where $\tau$ is the relative time coordinate, and $c_{\ve{i},\sigma}(\tau)\equiv e^{iH\tau}c_{\ve{i},\sigma}e^{-iH\tau}$. The $p$-wave opposite- and equal-spin triplet amplitudes are defined as 
$P_{0,\ve{i}}^{n}\equiv\sum_{\sigma}\left(\left<c_{\ve{i},\sigma}c_{\ve{i}+\ve{n},-\sigma}\right>-\left<c_{\ve{i},\sigma}c_{\ve{i}-\ve{n},-\sigma}\right>\right)$, and
$P_{\sigma,\ve{i}}^{n}\equiv\left<c_{\ve{i},\sigma}c_{\ve{i}+\ve{n},\sigma}\right>-\left<c_{\ve{i},\sigma}c_{\ve{i}-\ve{n},\sigma}\right>$, where $n\in\{x,y\}$. These are projected along the $z$ axis, but can be rotated to any projection axis.
The superconducting critical temperature $T_c$ is calculated by a binomial search where we for each temperature decide whether the gap has increased toward a superconducting state or decreased toward a normal state after a set number of iterative recalculations starting at an initial guess much smaller than the zero-temperature superconducting gap.
The free energy is given by $F=-T\:\text{ln}[\text{Tr}(e^{- H/T})]$ and is calculated using the eigenenergies of the system for a given temperature.

\begin{figure}[t]
    \centering
    \includegraphics[width=\columnwidth]{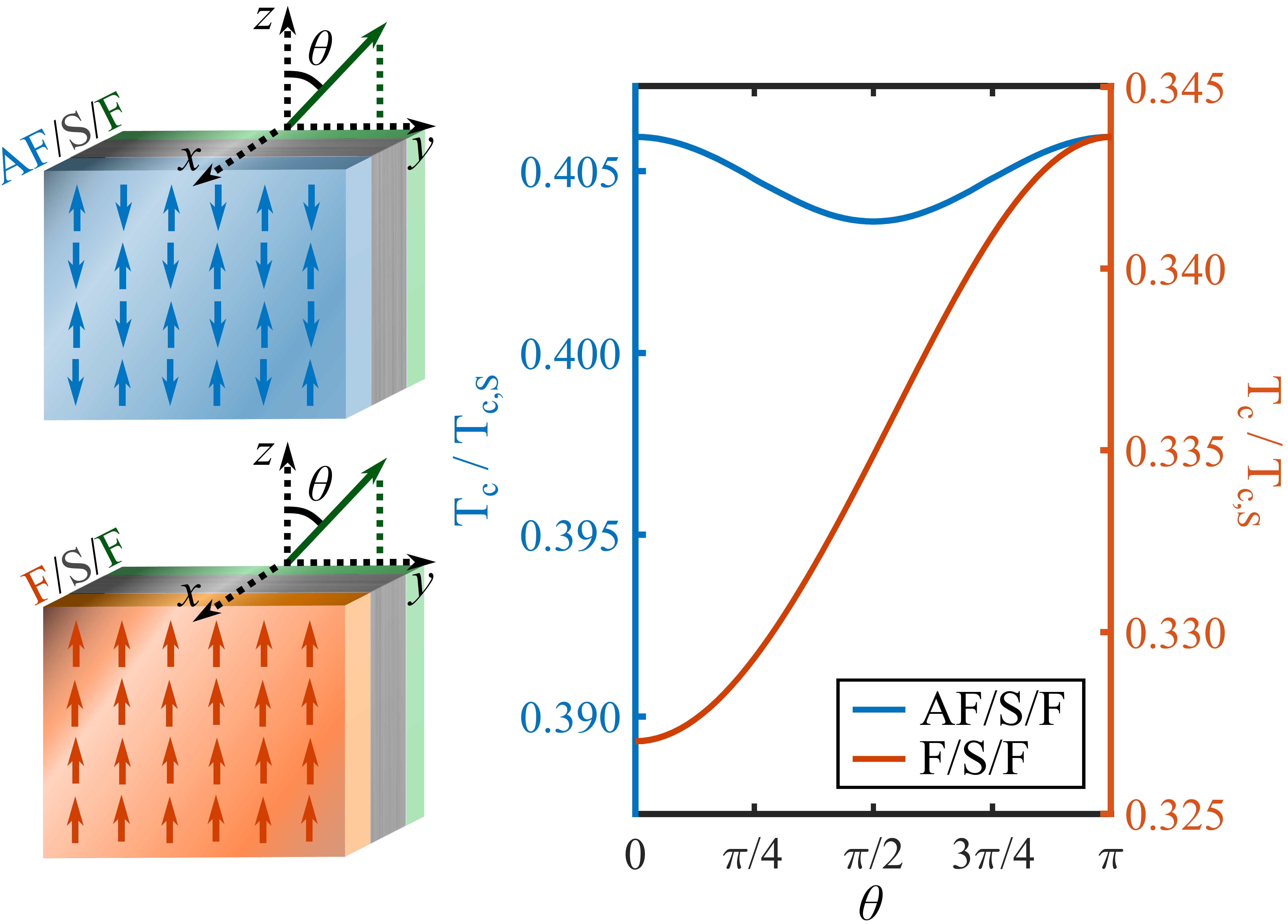}
    \caption{In the AF/S/F structure, we find a suppression of $T_c$ at $\theta=\pi/2$, when the magnetization of the ferromagnet is perpendicular to the magnetic moments of the antiferromagnet (blue curve). This variation in $T_c$ is only about seven times smaller than the difference in $T_c$ between antiparallel and parallel alignment of the magnetizations of the ferromagnets in a F/S/F structure (orange curve). In the plot, we have compared $T_c$ to the superconducting critical temperature without proximity to the magnetic layers, $T_{c,\text{S}}$. The parameters chosen for the AF/S/F system are $N_x^{\text{AF}}=4$, $N_x^{\text{S}}=9$, $N_x^{\text{F}}=3$, $N_y =90$, $t=1$, $\mu^{\text{AF}}=0.0001$, $\mu^{\text{S}}=\mu^{\text{F}}=0.9$, $M=0.4$, $U=2$, and $h=1$.
    The coherence length is comparable to the thickness of the superconducting layer. Qualitatively similar behavior in $T_c$ is found also for other choices of parameters. For the F/S/F structure, we replace the antiferromagnet with a ferromagnet with the same chemical potential as the rest of the structure and with magnetic exchange field of magnitude $h=M=0.4$ along the $z$ axis.
    }
    \label{fig:Tc}
\end{figure}

\textit{The superconducting critical temperature}.---
We first consider how the superconducting critical temperature of an AF/S/F hybrid structure with a compensated antiferromagnetic interface varies for in-plane rotations of the ferromagnetic magnetization.
As shown in Fig.~\ref{fig:Tc}, we find that $T_c$ decreases as the magnetization of the ferromagnet and the magnetic moments of the antiferromagnet are increasingly misaligned. Compared to a system where the antiferromagnetic layer is replaced by a ferromagnetic layer of the same thickness and with a magnetic exchange field of magnitude $h=M$, we find that the change in $T_c$ in the AF/S/F structure is only about seven times smaller. For F/S/F hybrids, experiments have demonstrated a difference in $T_c$ between parallel and antiparallel states of several hundreds of milli-Kelvin \cite{gu_prl_15}. This means that the difference in $T_c$ between aligned and perpendicular magnetic moments for an AF/S/F structure with a compensated interface should be measurable.

To understand why an antiferromagnet with a compensated interface, where the net magnetization is zero both in the bulk and at the interface, can be used to control the superconducting condensate, we first consider the more thoroughly studied F/S/F structure. In the F/S/F structure, we have two competing effects that determine $T_c$.
The dominant effect is the partial mutual compensation of the ferromagnetic exchange fields when the ferromagnets have antiparallel components \cite{leksin_prl_12}. For parallel alignment, the total magnetic field of the ferromagnets is stronger and superconductivity is more suppressed. This causes the variation in $T_c$ for the F/S/F structure seen in Fig.~\ref{fig:Tc}. 
The second weaker contribution to the $T_c$ variation is caused by spin-triplet generation that depends on the misalignment between the magnetic exchange fields of the two ferromagnets \cite{leksin_prl_12}. When the magnetizations of the two ferromagnets are misaligned, opposite-spin $s$- and $p_x$-wave triplets generated at one ferromagnetic interface are partly seen as equal-spin triplets with respect to the magnetization of the other ferromagnet. These have a much longer decay length inside the ferromagnet (up to hundreds of nm) compared to opposite-spin triplets which decay over a short length scale (of order nm). This opening of the equal-spin triplet channels causes a stronger suppression of $T_c$ when the ferromagnets are perpendicular. 

In our compensated AF/S/F system, the magnetic field from the antiferromagnet is zero, and the total magnetic field suppressing superconductivity is thus invariant under inversion of the magnetization of the ferromagnet. The dominant effect on $T_c$ in F/S/F structures is therefore absent in the compensated AF/S/F structure.
The triplet generation in F/S/F structures only depends on how much the magnetic moments of the two ferromagnets are misaligned, and not on whether they are parallel or antiparallel. However, when inverting the magnetization of one of the ferromagnets, the equal-spin triplet amplitudes with respect to the other ferromagnet changes sign. This means that the amplitude of long-range triplets in the ferromagnet is zero when we average over all up and down spins in the antiferromagnet. 
On the other hand, when the ferromagnetic magnetization is misaligned with the magnetic moments of the antiferromagnet, there is a finite equal-spin triplet amplitude with respect to the axis along which the magnetic moments of the antiferromagnet are aligned. We find these to be more robust to pair-breaking effects caused by the local magnetic exchange fields associated with the magnetic moments in the antiferromagnet. In contrast, spin-singlet and opposite-spin triplet Cooper pairs are more easily destroyed at the interface of the antiferromagnetic insulator. This is most likely caused by spin-up and spin-down electrons acquiring a $\pi$ phase difference upon reflection \cite{bobkova_prl_05}. At perpendicular alignment between the magnetic moments of the antiferromagnet and ferromagnet, the amplitude of the equal-spin triplets is at its maximum. In this case, more triplets are generated, causing a weakening of the superconducting condensate as more singlets are converted. This causes the $T_c$ variation in the compensated AF/S/F structure seen in Fig.~\ref{fig:Tc}. Plots showing the triplet amplitudes and superconducting gap are presented in the Supplemental Material \cite{Supplemental}.

Since the changes in $T_c$ only depend on the misalignment between the magnetic moments of the ferromagnet and antiferromagnet, not on their orientation with respect to the interface, we can choose to rotate the magnetization within the plane of the interface. This way, no components of the magnetization are perpendicular to the superconducting layer, and we thus avoid the appearance of demagnetizing currents close to the interface, as well as vortex formation. An in-plane magnetization is favored as long as the shape anisotropy of the ferromagnet is sufficiently strong. The $T_c$ variation in the present AF/S/F structure is partially caused by a variation in the $s$-wave triplet amplitude. We therefore expect the predicted $T_c$ modulation to be robust to impurity scattering and observable in diffusive systems as well as the ballistic limit systems covered by our theoretical framework. 

\begin{figure}[t]
    \centering
    \includegraphics[width=\columnwidth]{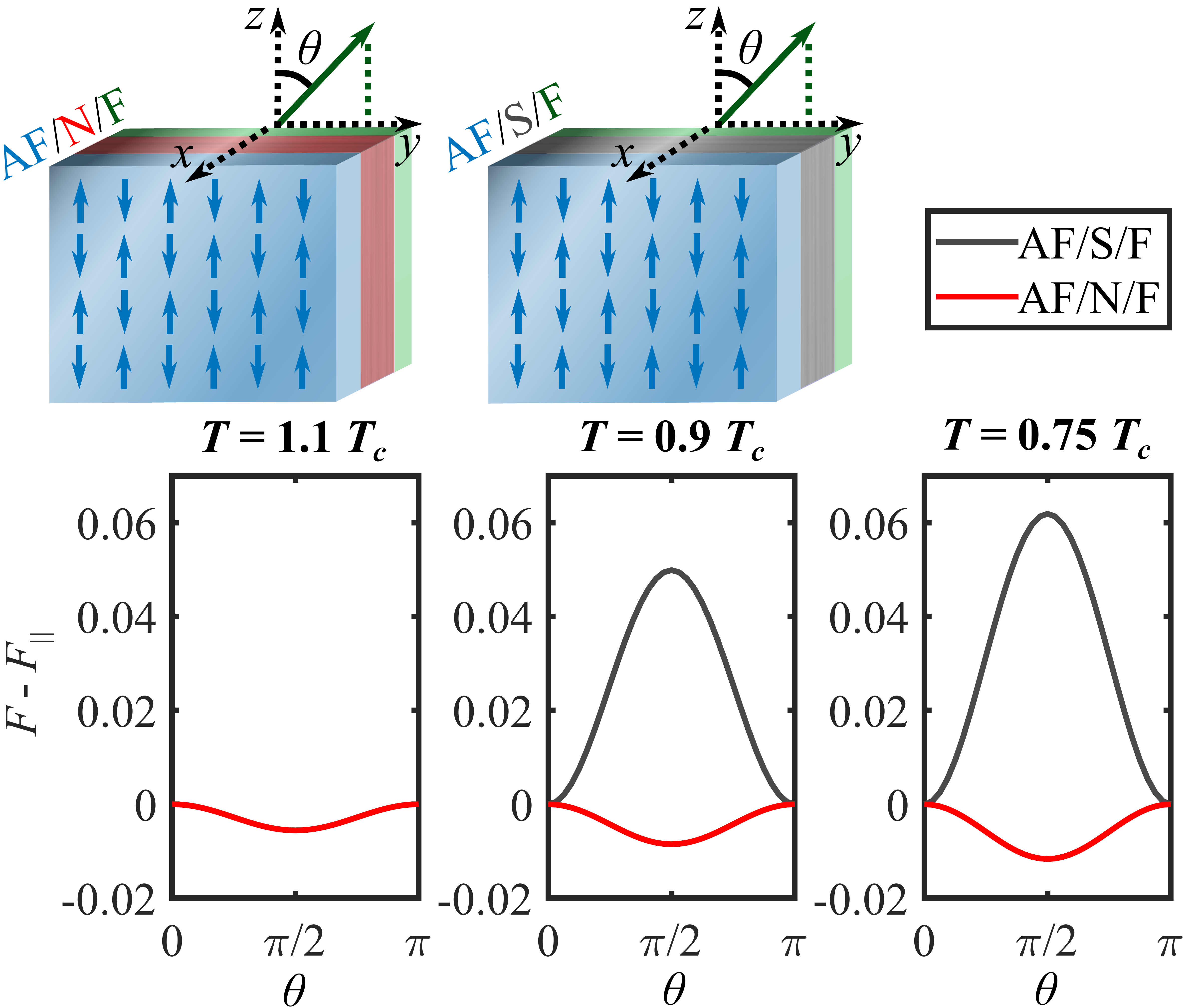}
    \caption{When decreasing the temperature below $T_c$, a peak develops in the free energy $F$ for a magnetization perpendicular to the magnetic moments of the antiferromagnet (grey curve). This causes a shift in the free energy minimum compared to the normal state (red curve) allowing for a $\pi/2$ in-plane rotation of the magnetization. The free energy is plotted relative to the free energy $F_{||}$ for parallel alignment for easier comparison between the normal-state and superconducting free energy. The parameters used are $N_x^{\text{AF}}=4$, $N_x^{\text{S}}=12$, $N_x^{\text{F}}=3$, $N_y =60$, $t=1$, $\mu^{\text{AF}}=0.0001$, $\mu^{\text{S}}=\mu^{\text{F}}=0.9$, $M=0.5$, $U=1.7$, and $h=0.7$. This corresponds to a coherence length comparable to the thickness of the superconducting region.
    }
    \label{fig:F}
\end{figure}

\textit{Magnetization reorientation}.---
Until now, we have explained how we can control the triplet channels in a compensated AF/S/F structure in order to manipulate the superconducting critical temperature. We now investigate another consequence of the weakening of the superconducting condensate, namely an increase in the free energy. Since the superconducting condensate is at its weakest for perpendicular alignment of the magnetic moments of the ferromagnet and antiferromagnet, we expect the superconducting contribution to the free energy to be at its maximum. If a perpendicular orientation is preferred for temperatures above $T_c$, we can achieve a rotation of the ground state magnetization direction by decreasing the temperature below $T_c$, as shown in Fig.~\ref{fig:F}. Assuming that the shape anisotropy of the ferromagnet enforces in-plane magnetization, it is thus possible to have $\pi/2$ magnetization reorientation within the plane of the interface driven by the superconducting phase transition. 
Similar predictions for S/F structures with interfacial spin-orbit coupling \cite{johnsen_prb_19} have been supported by experiments \cite{gonzalez-ruano_prb_20}. 

The normal-state free energy shown in Fig.~\ref{fig:F} only gives an example of a possible normal-state free energy curve for the compensated AF/S/F system. For experimentally realizing the magnetization reorientation, one must ensure that the magnetization in the normal-state is not aligned with the magnetic moments of the antiferromagnet. Using our Bogoliubov--de~Gennes theoretical framework, the normal-state free energy depends strongly on the choice of parameters. The exaggerated variation in the normal-state free energy under rotations of the magnetization is a thermal effect caused by an overestimated critical temperature, for the following reason. When considering a Bogoliubov--de Gennes Hamiltonian for a lattice structure, the lattice needs to be scaled down for the system to be computationally manageable. Since the superconducting coherence length is inversely proportional to the zero-temperature superconducting gap, we need the superconducting gap and thus $T_c$ to be large in order to have a coherence length comparable to the thickness of the superconductor. However, it is only the normal-state free energy that is substantially affected by the high temperatures. This is because the temperature dependence of the free energy strongly depends on the eigenenergies close to zero energy \cite{johnsen_prb_19}. In the presence of a superconducting gap, few eigenenergies exist in this range. We have chosen our parameters so that the coherence length is comparable to the thickness of the superconductor, and the magnetic exchange field is about one order of magnitude larger than the superconducting gap. Predictions based on similar scaling have previously corresponded well to experiments (see \eg Refs.~\cite{black-schaffer_prb_10,english_prb_16} and Refs.~\cite{johnsen_prb_19,banerjee_prb_18,gonzalez-ruano_prb_20}).

\textit{Concluding remarks}.---
In this letter, we have shown that the misalignment between the magnetic moments of an antiferromagnet and a ferromagnet is sufficient for controlling the triplet channels in an AF/S/F heterostructure, even when the antiferromagnetic interface is compensated and thus has zero effective interfacial magnetization. 
This provides the possibility of tuning the superconducting critical temperature by rotating the magnetization of a single ferromagnetic layer within the plane of the interface. In this way, the superconducting condensate can easily be controlled without having to deal with multiple ferromagnetic regions or out-of-plane magnetic fields causing demagnetizing currents and vortex formation in the superconducting region. 
Furthermore, we find that the superconducting transition can trigger a $\pi/2$ rotation of the ferromagnetic magnetization within the plane of the interface, allowing for temperature-controlled magnetic switching.

\begin{acknowledgments}
We thank V. Risinggaard, A. Kamra, and I. Bobkova for useful discussions. We acknowledge funding via the ``Outstanding Academic Fellows'' programme at NTNU, the Research Council of Norway Grant numbers 302315, as well as through its Centres of Excellence funding scheme, project number 262633, “QuSpin”.
\end{acknowledgments}

\section{Supplemental Material}

We here provide a more detailed description of our theoretical framework (Section~\ref{sec:theory}), where we describe the mean-field treatment and diagonalization of the Hamiltonian, as well as how we calculate the superconducting gap, coherence length, critical temperature, and the free energy. We use a similar approach as in Refs.~\cite{andersen_prb_05,leksin_prl_12,johnsen_prb_19,johnsen_prl_20}. We also provide further discussion of the triplet amplitudes present in the AF/S/F structure (Section~\ref{sec:triplets}).

\subsection{Theoretical framework}
\label{sec:theory}

We first use the mean-field approximation described in the main text on the antiferromagnetic and superconducting terms in the Hamiltonian given in Eq.~(1) in the Letter. The antiferromagnetic contribution to the Hamiltonian takes the form
\begin{equation}
H_{\text{AF}}=\frac{3}{8}\sum_{\boldsymbol{i}}\frac{M_{\boldsymbol{i}}^2}{V_{\boldsymbol{i}}}
-\sum_{\boldsymbol{i}} M_{\boldsymbol{i}}(n_{\boldsymbol{i},\uparrow}-n_{\boldsymbol{i},\downarrow})
\label{Eqn:H_AF}
\end{equation}
while the superconducting contribution takes the form
\begin{equation}
H_{\text{S}}=
\sum_{\boldsymbol{i}}{\frac{|\Delta_{\boldsymbol{i}}|}{U_{\boldsymbol{i}}}}+ \sum_{\boldsymbol{i}}{(\Delta_{\boldsymbol{i}}c^\dagger_{\boldsymbol{i}\uparrow}c^\dagger_{\boldsymbol{i}\downarrow}+\textrm{h.c.})}.
\label{Eqn:H_S}
\end{equation}
The superconducting gap $\Delta_{\ve{i}}$ is calculated self-consistently. The absolute value of antiferromagnetic order parameter $M_{\ve{i}}$ is assumed to be constant inside the antiferromagnetic region, while the sign alternates between neighboring lattice sites.

Bogoliubov--de Gennes lattice models are typically simplified by Fourier transforming in directions other than along the junction normal. For our $2$D lattice we therefore apply periodic boundary conditions in the $y$ direction. The two-sublattice periodic ordering of antiferromagnets has oscillating magnetic order parameter. This means we can still perform the Fourier transform if we take account of the doubling of the magnetic period \cite{andersen_thesis_04,andersen_prb_05}. In effect, this means that the magnetic order parameter must have a $\left\{\boldsymbol{k},\boldsymbol{k}+\boldsymbol{Q}\right\}$ symmetry for a reciprocal lattice vector $2\boldsymbol{Q}$, doubling the size of the matrix space.
The general expression for the Fourier transform in the $y$ direction is given by
\begin{eqnarray}
c_{\boldsymbol{i},\sigma}=\frac{1}{\sqrt{N_y}}\sum_{k_y}{c_{i_x ,k_y ,\sigma}e^{ik_yi_y}}.
\label{Eqn:FT}
\end{eqnarray}
When Fourier transforming the antiferromagnetic term, we take account of the oscillation at every lattice point explicitly by including $(-1)^{i_y}\equiv e^{2\pi i(i_y)}$, which gives the $\left\{\boldsymbol{k},\boldsymbol{k}+\boldsymbol{Q}\right\}$ symmetry when using
\begin{equation}
\begin{split}
    \label{rel}
    &\frac{1}{\sqrt{N_y}}\sum_{i_y} e^{i(k_y-k_y ')i_y}=\delta_{k_y , k_y '}.
\end{split}
\end{equation} 

After Fourier transforming, we write the Hamiltonian as
\begin{eqnarray}
H=H_0+\frac{1}{2}\sum_{i_x ,j_x ,k_y}{B^\dagger_{i_x ,k_y} H_{i_x ,j_x ,k_y} B_{j_x k_y}},
\label{Eqn:H_B}
\end{eqnarray}
where we have introduced the basis
\begin{eqnarray}
B^\dagger_{i_x ,k_y}&&=\left[c^\dagger_{i_x ,k_y ,\uparrow} \hspace{0.2cm} c^\dagger_{i_x ,k_y ,\downarrow} \hspace{0.2cm} c^\dagger_{i_x ,k_y +Q ,\uparrow} \hspace{0.2cm} c^\dagger_{i_x ,k_y +Q,\downarrow}\right.\nonumber\\
&& \left.\hspace{0.42cm} c_{i_x ,-k_y ,\uparrow} \hspace{0.2cm} c_{i_x ,-k_y ,\downarrow} \hspace{0.2cm} c_{i_x ,-k_y -Q,\uparrow} \hspace{0.2cm} c_{i_x ,-k_y -Q,\downarrow}  \right].
\label{Eqn:Basis}
\end{eqnarray}
The matrix $H_{i_x ,j_x ,k_y}$ is given by
\begin{eqnarray}
H_{i_x ,jx ,k_y}&=&-\left[\frac t2(\delta_{i_x ,j_x -1}+\delta_{i_x ,j_x +1})+\frac{\mu}{2}\delta_{i_x ,j_x}\right]\hat{\tau}_3\hat{\rho}_0\hat{\sigma}_0\nonumber\\
&&-t\cos(k_y )\delta_{i_x ,j_x }\hat{\tau}_3\hat{\rho}_3\hat{\sigma}_0 -\frac 12 M_{i_x}\delta_{i_x ,j_x }\hat{\tau}_3\hat{\rho}_1\hat{\sigma}_3\nonumber\\
&&+\frac{\delta_{i_x ,j_x }}{2}\left(h_{i_x} ^x\hat{\tau}_3\hat{\rho}_0\hat{\sigma}_1+h_{i_x}^y\hat{\tau}_0\hat{\rho}_0\hat{\sigma}_2+h_{i_x}^z\hat{\tau}_3\hat{\rho}_0\hat{\sigma}_3\right)\nonumber\\
&&+\frac i2\delta_{i_x ,j_x }\Delta_{i_x}\hat{\tau}^+\hat{\rho}_0\hat{\sigma}_2-\frac i2 \delta_{i_x ,j_x }\Delta_{i_x}^*\hat{\tau}^-\hat{\rho}_0\hat{\sigma}_2.
\label{Eqn:H_ijk}
\end{eqnarray}
Here the matrices $\hat{\tau}_i$, $\hat{\rho}_i$ and $\hat{\sigma}_i$ for $i=\{0,1,2,3\}$ are the usual $SU(2)$ (Pauli) matrices, with $i=0$ being the identity, and with $\hat{\tau}^\pm =(\hat{\tau}_1 \pm i\hat{\tau}_2 )/2$. $\hat{\tau}$ represents particle-hole space, $\hat{\rho}$ the $\left\{\boldsymbol{k},\boldsymbol{k}+\boldsymbol{Q}\right\}$-space and $\hat{\sigma}$ denotes spin space, as can be identified from the basis choice. 
The constant term $H_0$ is given by
\begin{eqnarray}
H_0&=&-\frac{1}{2N_y}\sum_{i_x}\mu_{i_x}-\sum_{i_x ,k_y}t\cos(k_y )\nonumber\\
&&+\sum_{i_x}{\frac{N_y|\Delta_{i_x }|^2}{U_{i_x}}}+\frac{3}{8}\sum_{i_x}\!{\frac{N_y M^2_{i_x}}{V_{i_x}}}.
\label{Eqn:H0}
\end{eqnarray}
By defining another basis,
\begin{equation}
    W_{k_y , k_z}^\dagger = [B_{1,k_y }^\dagger,...,B_{i_x ,k_y }^\dagger ,...,B_{N_x ,k_y }^\dagger ],
\end{equation}
Eq. \ref{Eqn:H_B} can be rewritten as 
\begin{equation}
    H=H_0 + \frac{1}{2}\sum_{k_y}W_{k_y}^\dagger H_{k_y } W_{k_y},
\end{equation}
where
\begin{equation}
\label{Hkykz}
    H_{k_y}=
    \begin{bmatrix}
        H_{1,1,k_y } & \cdots & H_{1,N_x ,k_y }\\
        \vdots & \ddots & \vdots \\
        H_{N_x ,1,k_y } & \cdots & H_{N_x ,N_x ,k_y }\\
    \end{bmatrix}.
\end{equation}
We diagonalize $H_{k_y }$ numerically and obtain eigenvalues $E_{n,k_y }$ and eigenvectors $\Phi_{n,k_y }$ given by 
\begin{equation}
\begin{split}
    \Phi_{n,k_y }\equiv[& \phi_{1,n ,k_y} \hspace{3mm} \cdots \hspace{3mm}\phi_{N_x ,n ,k_y}]^T ,\\
    \phi_{i_x ,n,k_y }\equiv[&s_{i_x ,n,k_y }\hspace{2mm} t_{i_x ,n,k_y }\hspace{2mm} u_{i_x ,n,k_y }\hspace{2mm} v_{i_x ,n,k_y }\\
    &w_{i_x ,n,k_y }\hspace{2mm} x_{i_x ,n,k_y }\hspace{2mm} y_{i_x ,n,k_y }\hspace{2mm} z_{i_x ,n,k_y }]^T .
\end{split}
\end{equation} 
In its diagonalized form the Hamiltonian can be written as
\begin{eqnarray}
H=H_0+\frac12\sum_{n,k_y}E_{n,k_y}\gamma^\dagger_{n,k_y}\gamma_{n,k_y},
\label{Eqn:Hgamma}
\end{eqnarray}
where the new quasiparticle fermion operators $\gamma_{n,k_y}$ satisfy
\begin{eqnarray}
c_{i_x ,k_y ,\uparrow} \!=\! \sum\limits_{n}{s_{i_x ,n,k_y}\gamma_{n,k_y}}, \hspace{0.7cm}c^\dagger_{i_x,-k_y,\uparrow} \!&=\! \sum\limits_{n}{w_{i_x ,n,k_y}\gamma_{n,k_y}},\nonumber\\
c_{i_x ,k_y ,\downarrow} \!=\! \sum\limits_{n}{t_{i_x ,n,k_y}\gamma_{n,k_y}}, \hspace{0.7cm}c^\dagger_{i_x ,-k_y ,  \downarrow} \!&=\! \sum\limits_{n}{x_{i_x ,n,k_y}\gamma_{n,k_y}},\nonumber\\
c_{i_x ,k_y +Q,\uparrow} \!=\! \sum\limits_{n}{u_{i_x ,n,k_y }\gamma_{n,k_y }}, \hspace{0.2cm}c^\dagger_{i_x ,-k_y -Q,\uparrow} \!&=\! \sum\limits_{n}{y_{i_x ,n,k_y}\gamma_{n,k_y}},\nonumber\\
c_{i_x ,k_y +Q,\downarrow} \!=\! \sum\limits_{n}{v_{i_x ,n,k_y}\gamma_{n,k_y}}, \hspace{0.2cm}c^\dagger_{i_x ,-k_y -Q,\downarrow} \!&=\! \sum\limits_{n}{z_{i_x ,n,k_y}\gamma_{n,k_y}}.\nonumber\\
\label{Eqn:gamma}
\end{eqnarray}
The Fermi-Dirac distribution function $f$ allows us to calculate expectation values of the form $\langle \gamma^\dagger_{n,k_y}\gamma_{m,k_y}\rangle=f(E_{n,k_y}/2)\delta_{n,m}$.

\begin{figure}[bt]
    \centering
    \includegraphics[width=\columnwidth]{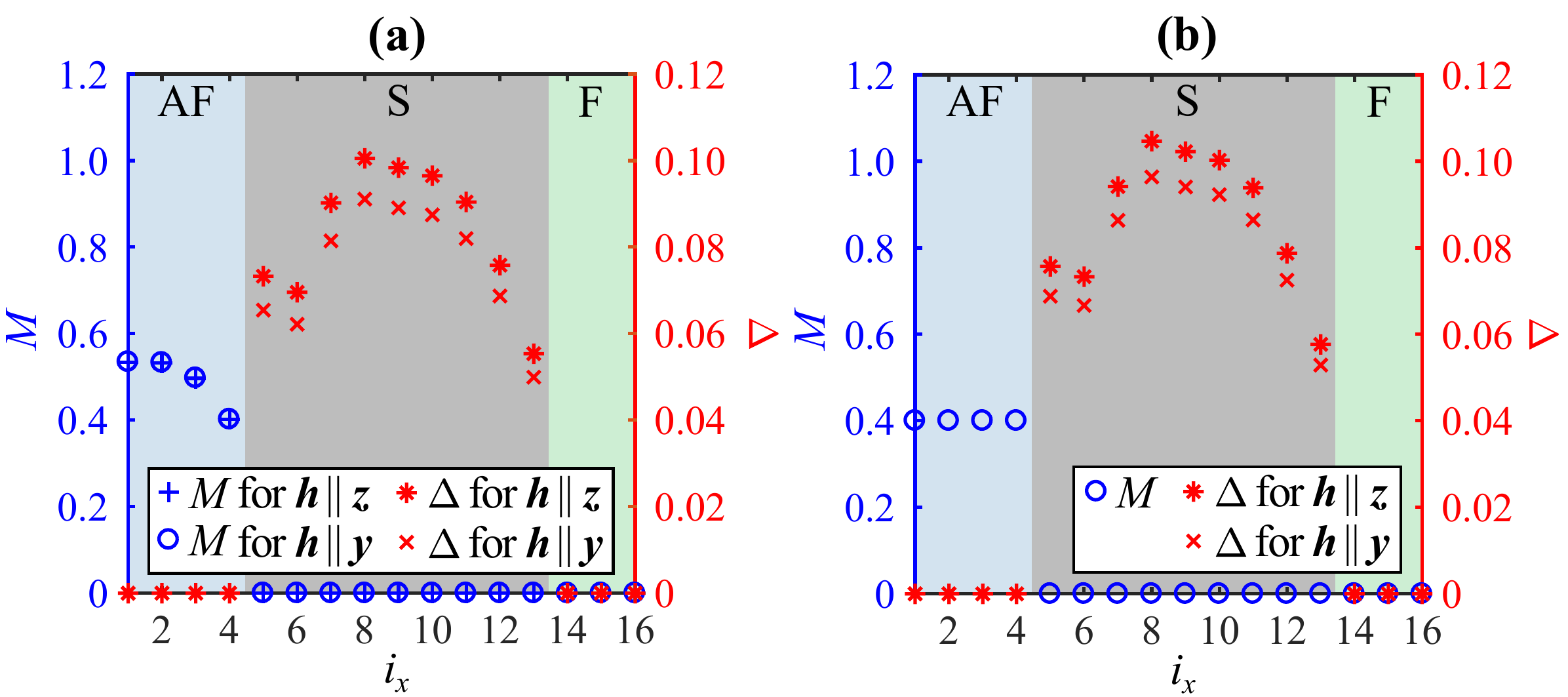}
    \caption{Panel (a) and (b) show the magnitude of the antiferromagnetic order parameter $M$ and the superconducting gap $\Delta$ as a function of the position $i_x$ at temperature $T_c /2$ for the same parameters as used in Fig.~2 in the manuscript. In panel (a) both $M$ and $\Delta$ are calculated self-consistently with $V=2.35$. We see that when rotating the magnetization of the ferromagnet from a parallel ($\boldsymbol{h}||\boldsymbol{z}$) to a perpendicular ($\boldsymbol{h}||\boldsymbol{y}$) alignment with respect to the magnetic moments of the antiferromagnet, the superconducting gap changes significantly, while the antiferromagnetic order parameter is unchanged. In panel (b), only $\Delta$ is solved self-consistently, as in our manuscript. The order parameter of the antiferromagnet is set equal to its smallest value in the self-consistent solution. We find a similar variation in $\Delta$ as in panel (a). }
    \label{fig:self-consistent_AF_OP}
\end{figure}

\begin{figure*}[tb]
    \centering
    \includegraphics[width=\textwidth]{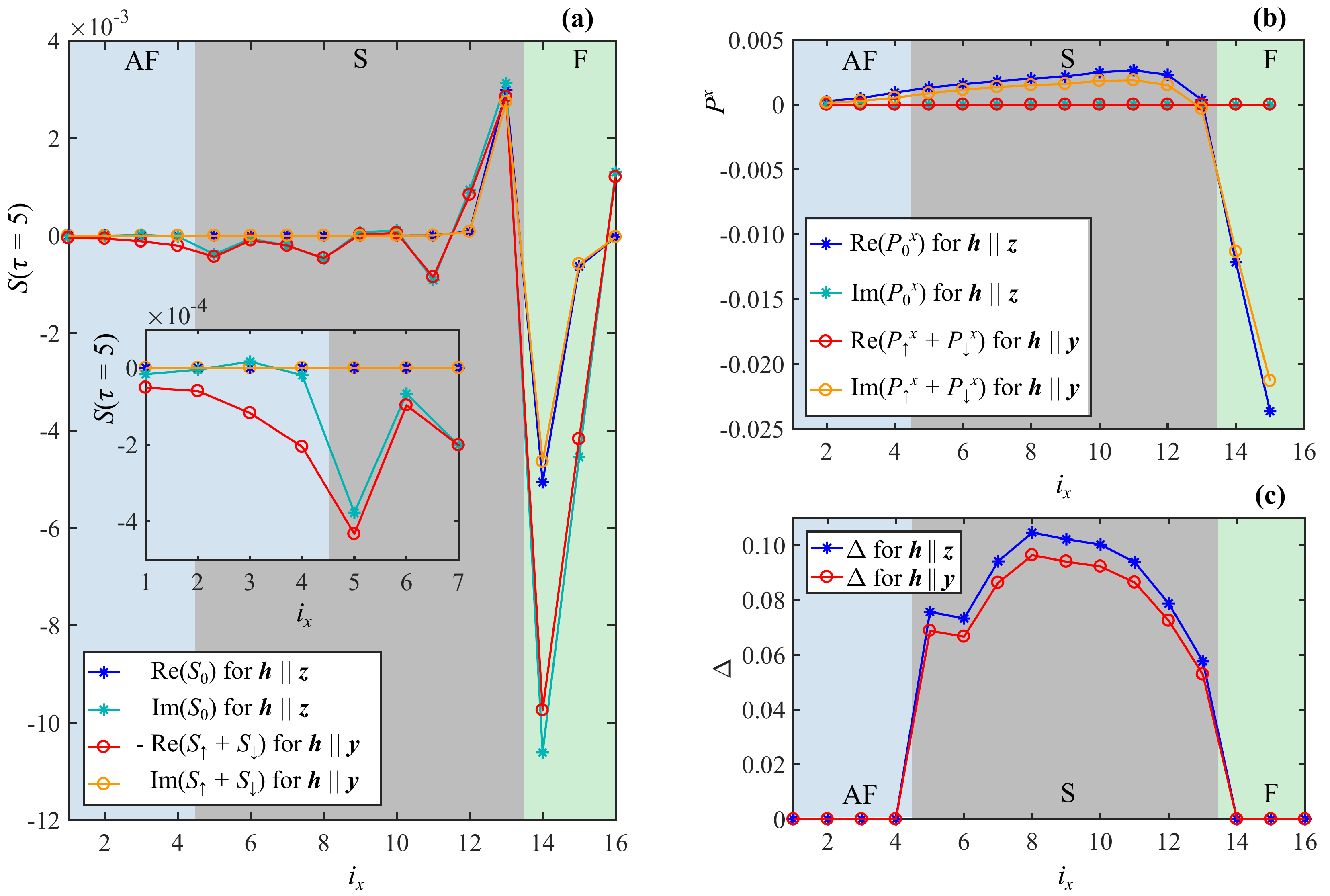}
    \caption{Panels (a)-(c) show the $s$-wave triplet amplitude, the $p_x$-wave triplet amplitude and the superconducting gap, respectively, at temperature~$T_c /2$ inside the antiferromagnetic (blue), superconducting (grey) and ferromagnetic (green) region. In all panels we compare the results for parallel ($\boldsymbol{h}||\boldsymbol{z}$) and perpendicular ($\boldsymbol{h}||\boldsymbol{y}$) alignment of the magnetic moments of the antiferromagnet and ferromagnet. The inset in panel~(a) shows the $s$-wave triplet amplitudes inside and close to the antiferromagnetic region. The parameters used are the same as for Fig.~2 in the main text.}
    \label{fig:triplets}
\end{figure*}

The superconducting gap is defined $\Delta_{\ve{i}}\equiv U_{\ve{i}}\left<c_{\ve{i},\uparrow}c_{\ve{i},\downarrow}\right>$, and can be expressed in terms of the eigenenergies and elements of the eigenvectors as 
\begin{equation}
    \begin{split}
\Delta_{i_x}=&-\frac{U_{i_x}}{2N_y}\sum_{n,k_y}(t_{i_x ,n,k_y}w^*_{i_x ,n,k_y }\\
&+v_{i_x ,n,k_y }y^*_{i_x ,n,k_y})[1- f(E_{n,k_y}/2)].
    \end{split}
    \label{Eqn:SCOP}
\end{equation}
Similarly, the $s$-wave odd-frequency opposite- and equal-spin triplet amplitudes can be written
\begin{equation}
    \begin{split}
        S_{0,i_x}(\tau)=&\frac{1}{2 N_y }\sum_{n,k_y }[s_{i_x ,n,k_y }x_{i_x ,n,k_y }^{*}
        +t_{i_x ,n,k_y }w_{i_x ,n,k_y }^{*}\\
        &+u_{i_x ,n,k_y }z_{i_x ,n,k_y }^{*}
        +v_{i_x ,n,k_y }y_{i_x ,n,k_y }^{*}]
        e^{-iE_{n,k_y }\tau/2} \\
        &\cdot[1-f(E_{n,k_y }/2)],\\
        S_{\uparrow,i_x}(\tau)=&\frac{1}{2N_y }\sum_{n,k_y }[s_{i_x ,n,k_y }w_{i_x ,n,k_y }^{*}+u_{i_x ,n,k_y }y_{i_x ,n,k_y }^{*}]\\
        &\cdot e^{-iE_{n,k_y }\tau/2}[1-f(E_{n,k_y }/2)],\\
        S_{\downarrow,i_x}(\tau)=&\frac{1}{2N_y }\sum_{n,k_y }[t_{i_x ,n,k_y }x_{i_x ,n,k_y }^{*}+v_{i_x ,n,k_y }z_{i_x ,n,k_y }^{*}]\\
        &\cdot e^{-iE_{n,k_y }\tau/2}[1-f(E_{n,k_y }/2)],
    \end{split}
\end{equation}
and the $p_x$-wave even-frequency opposite- and equal-spin triplet amplitudes are given by
\begin{equation}
    \begin{split}
        P_{0,i_x}^{x}=&\frac{1}{2N_y}\sum_{n,k_y }[s_{i_x ,n,k_y }x_{i_x +1 ,n,k_y }^{*}-s_{i_x ,n,k_y }x_{i_x -1 ,n,k_y }^{*}\\
        &+u_{i_x ,n,k_y }z_{i_x +1 ,n,k_y }^{*}-u_{i_x ,n,k_y }z_{i_x -1 ,n,k_y }^{*}\\
        &+t_{i_x ,n,k_y }w_{i_x +1 ,n,k_y }^{*}-t_{i_x ,n,k_y }w_{i_x -1 ,n,k_y }^{*}\\
        &+v_{i_x ,n,k_y }y_{i_x +1 ,n,k_y }^{*}-v_{i_x ,n,k_y }y_{i_x -1 ,n,k_y }^{*}]\\
        &\cdot[1-f(E_{n,k_y }/2)],\\
        P_{\uparrow,i_x}^{x}=&\frac{1}{2N_y }\sum_{n,k_y }[s_{i_x ,n,k_y }w_{i_x +1 ,n,k_y }^{*}-s_{i_x ,n,k_y }w_{i_x -1 ,n,k_y }^{*}\\
        &+u_{i_x ,n,k_y }y_{i_x +1 ,n,k_y }^{*}-u_{i_x ,n,k_y }y_{i_x -1 ,n,k_y }^{*}]\\
        &\cdot[1-f(E_{n,k_y }/2)],\\
        P_{\downarrow,i_x}^{x}=&\frac{1}{2N_y }\sum_{n,k_y }[t_{i_x ,n,k_y }x_{i_x +1 ,n,k_y }^{*}-t_{i_x ,n,k_y }x_{i_x -1 ,n,k_y }^{*}\\
        &+v_{i_x ,n,k_y }z_{i_x +1 ,n,k_y }^{*}-v_{i_x ,n,k_y }z_{i_x -1 ,n,k_y }^{*}]\\
        &\cdot[1-f(E_{n,k_y }/2)].
    \end{split}
\end{equation}
Only $s$- and $p_x$-wave triplets are present in the AF/S/F structure. The above spin-triplet amplitudes describe spins projected along the $z$ axis. We rotate the triplets to a new projection axis characterized by the polar coordinate $\theta$ and the azimuthal angle $\phi$ with respect to the $z$ axis using
\begin{equation}
\label{S_new_basis}
    \begin{split}
        (\uparrow\downarrow+\downarrow\uparrow)_{\theta,\phi}=&-\sin(\theta)[e^{-i\phi}(\uparrow\uparrow)_z -e^{i\phi}(\downarrow\downarrow)_z ]\\
        &+\cos(\theta)(\uparrow\downarrow+\downarrow\uparrow)_z , \\
        (\uparrow\uparrow)_{\theta,\phi}=&\cos^2 (\theta/2)e^{-i\phi}(\uparrow\uparrow)_z +\sin^2 (\theta/2)e^{i\phi}(\downarrow\downarrow)_z\\
        &+\sin(\theta/2)\cos(\theta/2)(\uparrow\downarrow+\downarrow\uparrow)_z ,\\
        (\downarrow\downarrow)_{\theta,\phi}=&\sin^2 (\theta/2)e^{-i\phi}(\uparrow\uparrow)_z +\cos^2 (\theta/2)e^{i\phi}(\downarrow\downarrow)_z \\
        &-\sin(\theta/2)\cos(\theta/2)(\uparrow\downarrow+\downarrow\uparrow)_z ,
    \end{split}
\end{equation}
where $(\uparrow\downarrow+\downarrow\uparrow)$ represents the opposite-spin triplet amplitude, while $(\uparrow\uparrow)$ and $(\downarrow\downarrow)$ represents the equal-spin triplet amplitudes.

The antiferromagnetic order parameter can be calculated self-consistently from
\begin{align}
    &M_{i_x}=\frac{V}{4N_y}\sum_{n,k_y}[v_{i_x ,n,k_y}t^*_{i_x ,n,k_y} +t_{i_x ,n,k_y}v^*_{i_x ,n,k_y}\notag\\
    &-u_{i_x ,n,k_y}s^*_{i_x ,n,k_y}-s_{i_x ,n,k_y}u^*_{i_x ,n,k_y}][1-f(E_{n,k_y}/2)].
\end{align}
In Fig.~\ref{fig:self-consistent_AF_OP}, we justify our assumption that a self-consistent calculation of the antiferromagnetic order parameter is unnecessary. From panel~(a), where both the antiferromagnetic and superconducting order parameters are computed self-consistently, we see that the antiferromagnetic order parameter is suppressed close to the interface. However, as seen in panel~(b) where only the superconducting order parameter is computed self-consistently, setting the magnetic order parameter equal to the interfacial value of the the magnetic order parameter in panel~(a) gives practically identical results for the self-consistently obtained superconducting gap.

The superconducting coherence length is given by $\xi=\hbar v_F /\pi\Delta_0$, where $v_F\equiv\frac{1}{\hbar}\frac{dE_{k_y }}{dk}\big|_{k=k_F}$ is the normal-state Fermi velocity, $E_{k_y }$ is the normal-state eigenenergies if we instead of accounting for an interface use periodic boundary conditions along all three axes, and $k_F$ is the corresponding Fermi momentum averaged over the Fermi surface.

Our binomial search algorithm for the superconducting critical temperature, is the same as the one we have previously presented in Ref.~\cite{johnsen_prl_20}, and we only summarize the main steps here. We divide the temperature interval $N_T$ times. For each of the $N_T$ temperatures considered, we recalculate the gap $N_{\Delta}$ times from an initial guess with a magnitude much smaller than the zero-temperature superconducting gap. If the gap has increased towards a superconducting solution after $N_{\Delta}$ iterations, we conclude that the current temperature is below $T_c$. If the gap has decreased towards a normal-state solution, we conclude that the current temperature is above $T_c$.  The advantage of this algorithm is that we are not dependent upon recalculating the gap until it converges. The number of iterations $N_{\Delta}$ must only be large enough that we have an overall increase or decrease in the superconducting gap at all lattice sites inside the superconducting region under recalculation.

The minimum of the free energy defines the ground state of the system. The free energy is given by
\begin{equation}
    \label{F}
    F=H_0 -T\sum_{n,k_y }\ln(1+e^{- E_{n, k_y }/2T}).
\end{equation}
Note also that when $T\to 0$,
\begin{equation}
    \label{FT->0}
    F=H_0 +\frac{1}{2}\sum_{n,k_y} E_{n,k_y },
\end{equation}
where the sum is restricted to negative eigenenergies.

\subsection{Triplet amplitudes}
\label{sec:triplets}

In Fig.~\ref{fig:triplets}, we show the $s$-wave and $p_x$-wave triplet amplitudes and the superconducting gap inside the AF/S/F structure for parallel ($\boldsymbol{h}||\boldsymbol{z}$) and perpendicular ($\boldsymbol{h}||\boldsymbol{y}$) alignment of the magnetic moments inside the antiferromagnet and ferromagnet.
To understand how these triplets are generated, we need to comment on two limiting cases:
First, in the absence of the ferromagnetic layer all triplet amplitudes are zero. This means that all triplets in the AF/S/F structure are generated due to the proximity to the ferromagnet. 
Second, in the absence of the antiferromagnetic layer, all triplet amplitudes are invariant under rotation of the magnetization of the ferromagnet. This means that if we choose a projection axis along the $z$ axis, $S_0$ and $P_0^x$ for $\boldsymbol{h}||\boldsymbol{z}$ is equal to $S_{\uparrow}+S_{\downarrow}$ and $P^x_{\uparrow}+P^x_{\downarrow}$ for $\boldsymbol{h}||\boldsymbol{y}$, respectively. 
We now consider an AF/S/F structure where the magnetic moments of the antiferromagnet are directed along~$\pm\boldsymbol{z}$.
If we first consider the $s$-wave triplet amplitude (Fig.~\ref{fig:triplets}(a)), we find that the amplitude of the equal-spin triplets $S_{\uparrow}+S_{\downarrow}$ for $\boldsymbol{h}||\boldsymbol{y}$ is larger than the amplitude of the opposite-spin triplets $S_0$ for $\boldsymbol{h}||\boldsymbol{z}$ close to and inside the antiferromagnetic region (inset of Fig.~\ref{fig:triplets}(a)). While opposite-spin triplets hardly penetrate the antiferromagnet at all, the equal-spin triplets seem to have a longer penetration depth. Moreover, they seem to be more robust upon reflection. 
The robustness of the equal-spin triplets generated when the magnetic moments of the ferromagnet and antiferromagnet are misaligned explains the suppression of the superconducting gap (Fig.~\ref{fig:triplets}(c)) and superconducting critical temperature.
Close to and inside the ferromagnetic region, the $s$-wave triplet amplitude is instead weaker for $\boldsymbol{h}||\boldsymbol{y}$ than for $\boldsymbol{h}||\boldsymbol{z}$. This is because the triplet generation at the ferromagnetic interface is dependent of the scattering of triplets at the antiferromagnetic interface, via the singlet amplitude. The robustness of the equal-spin triplets close to and inside the antiferromagnet causes the superconducting gap (Fig.~\ref{fig:triplets}(c)) and thus the singlet amplitude to be smaller when $\boldsymbol{h}||\boldsymbol{z}$. The decreased singlet amplitude causes fewer triplets to be created by proximity to the ferromagnet. For the $s$-wave triplet amplitude this effect only dominates for positions sufficiently far away from the antiferromagnetic region, while for the $p_x$-wave triplet amplitude (Fig.~\ref{fig:triplets}(b)) it dominates in all regions.

\end{document}